\documentclass[aps,prb,amsmath,amsfonts,amssymb,notitlepage,twocolumn,superscriptaddress,footinbib,floatfix]{revtex4-2}

\usepackage{CJK}
\usepackage{bm}
\usepackage{soul}
\usepackage[caption=false,subrefformat=parens,labelformat=parens]{subfig}
\usepackage{color} 
\usepackage[table,dvipsnames]{xcolor}
\usepackage[colorlinks,linktocpage,bookmarks=false]{hyperref}

\usepackage{enumitem}
\usepackage{comment}
\usepackage[only,boxempty]{stmaryrd}
\usepackage{multirow}
\usepackage{xspace} 
\usepackage{tikz}
\newcommand\myshade{85}
\definecolor{myrulecolor}{RGB}{150,20,0}
\colorlet{mylinkcolor}{violet}
\colorlet{mycitecolor}{YellowOrange}
\colorlet{myurlcolor}{Aquamarine}
\hypersetup{
	linkcolor = myurlcolor!\myshade!black,
	citecolor = mycitecolor!\myshade!black,
	urlcolor = myrulecolor!\myshade!black,
}  

\usepackage{graphicx}
\graphicspath{{FIG/}}
\usepackage{multirow}
\usepackage{braket} 
\usepackage{booktabs}
\usepackage{ulem}

\AtBeginDocument{
	\heavyrulewidth=.08em
	\lightrulewidth=.05em
	\cmidrulewidth=.03em
	\belowrulesep=.65ex
	\belowbottomsep=0pt
	\aboverulesep=.4ex
	\abovetopsep=0pt
	\cmidrulesep=\doublerulesep
	\cmidrulekern=.5em
	\defaultaddspace=.5em
}

\newcommand{\beq}{\begin{equation}}
	\newcommand{\eeq}{\end{equation}}
\newcommand{\bea}{\begin{eqnarray}}
	\newcommand{\eea}{\end{eqnarray}}

\DeclareMathAlphabet\mathbfcal{OMS}{cmsy}{b}{n}

\newcommand\CP{\mathbb{C}P}
\newcommand\RP{\mathbb{R}P}

\newcommand{\bfR}{\mathbf{R}}

\newcommand{\calC}{\mathcal{C}}

\renewcommand\[{\begin{equation}}
	\renewcommand\]{\end{equation}}

\newcommand{\an}[1]{{\color{orange} [AN:] #1}}
\newcommand{\anc}[1]{{\footnotesize{\color{orange} [AN:] #1}}}

\definecolor{amethyst}{rgb}{0.6, 0.4, 0.8}

\makeindex

\makeatletter 
\newcommand*{\b@xplus}[1][+]{\ooalign{%
		$\m@th\vcenter{\hbox{$\m@th#1$}}$\cr%
		\hidewidth$\m@th\boxempty$\hidewidth\cr}} 
\renewcommand*{\boxplus}{\mathbin{\b@xplus}} 
\renewcommand*{\boxminus}{\mathbin{\b@xplus[-]}} 
\makeatother

\DeclareMathOperator{\rank}{rank}

\begin{document} 
	\begin{CJK*}{UTF8}{gbsn} 
		\title{Classification of Classical Spin Liquids:  Topological Quantum Chemistry and Crystalline Symmetry}
		\author{Yuan Fang}
		\affiliation{Department of Physics and Astronomy, Stony Brook University, Stony Brook, New York 11794, USA}	
		\author{Jennifer Cano}
		\affiliation{Department of Physics and Astronomy, Stony Brook University, Stony Brook, New York 11794, USA}
		\affiliation{Center for Computational Quantum Physics, Flatiron Institute, New York, New York 10010, USA}
		\author{Andriy H. Nevidomskyy} 
		\affiliation{Department of Physics and Astronomy, Rice University, Houston, TX 77005, USA}
		\affiliation{Rice Center for Quantum Materials, Rice University, Houston, TX 77005, USA}
		\author{Han Yan (闫寒)} 
		\affiliation{Department of Physics and Astronomy, Rice University, Houston, TX 77005, USA}
		\affiliation{Smalley-Curl Institute, Rice University, Houston, TX 77005, USA}
		\date{\today}
		\begin{abstract}
			Frustrated magnetic systems can host highly interesting phases known as classical spin liquids (CSLs), which feature  {extensive} ground state  degeneracy and lack long-range magnetic order.
			Recently,  Yan and Benton \textit{et al.}  proposed a classification scheme of CSLs in the large-$\mathcal{N}$ (soft spin) limit [\href{https://arxiv.org/abs/2305.00155}{arXiv:2305.00155}, \href{https://arxiv.org/abs/2305.19189}{arXiv:2305.19189}]. 
			This scheme classifies CSLs into two categories: the algebraic CSLs and the fragile topological CSLs, each with their own correlation properties, low energy effective description, and finer classification frameworks.  
			In this work, we further develop the classification scheme by considering the role of crystalline symmetry. 
			We present a mathematical framework for computing the band representation of the flat bands in the spectrum of these CSLs, which extends beyond the conventional representation analysis. 
			It allows one to determine whether the algebraic CSLs, which features gapless points on their bottom flat bands, are protected by symmetry or not. 
			It also provides more information on the finer classifications of algebraic and fragile topological CSLs. 
			We demonstrate this framework via concrete examples and showcase its power by constructing a pinch-line algebraic CSL protected by symmetry.
		\end{abstract}
		\maketitle
	\end{CJK*}  
	
	\section{Introduction}
	
	\begin{figure*}[th!]
		\centering
		\includegraphics[width=0.95\textwidth]{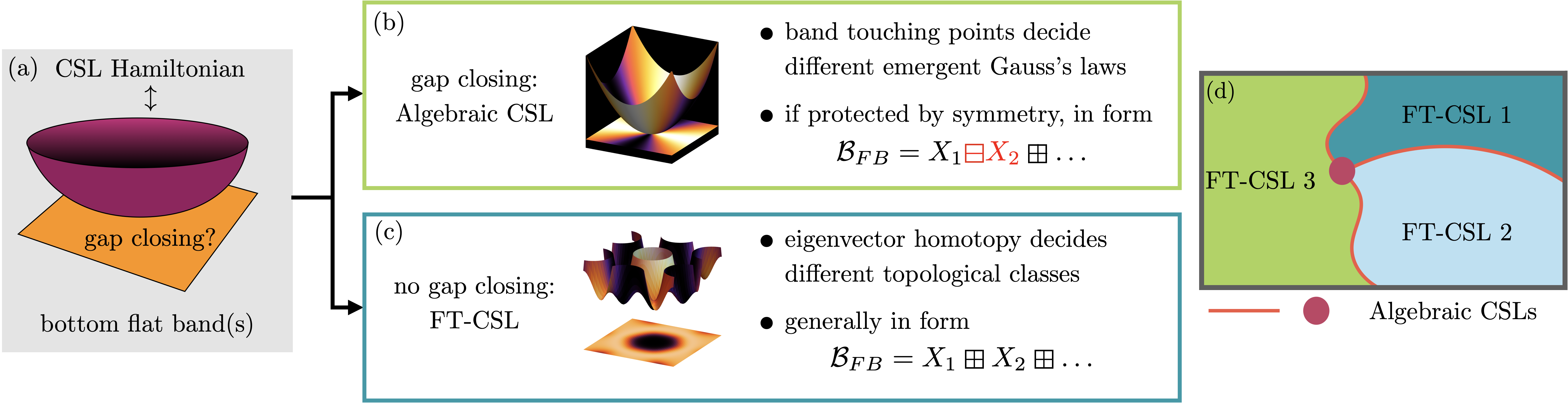} 
		\caption{
			(a) The CSL Hamiltonian generally features one or more degenerate flat bands at the bottom of its spectrum.
			(b) 
			Algebraic CSLs feature gap closing points between the   bottom flat bands and the dispersive bands. The band touching point determines the emergent Gauss's law. 
			$\mathcal{B}_{FB}$ is the flat band representation, and will be discussed in detail in the main text.
			(c) 
			Fragile Topological CSLs (FT-CSLs) have no such gap-closing points, and are classified by their eigenvector homotopy.
			(d) The landscape of the CSL phase diagram  consists of FT-CSLs whose boundaries are Algebraic CSLs.
		}
		\label{Fig_overview}
	\end{figure*}

	The investigation of magnetic systems lacking long-range order has a rich history, spanning several decades starting from the exploration of disorder effects in  spin glasses~\cite{Edwards1975,SK_glass} and the proposal of resonating valence bond states \cite{ANDERSON1973RVB,Anderson1987}, which have become fundamental to contemporary research in strongly frustrated magnets. A particular concept of interest is that of classical spin liquids (CSLs), which emerge when  spin models exhibit an extensively degenerate ground state manifold as a consequence of frustration, and fluctuations among ground states preclude any form of ordering \cite{Villain1979,MC_pyro_PRL,Moessner1998PhysRevB.58.12049,Isakov2004PhysRevLett.93.167204,Henley05PRB,Henley2010ARCMP,Garanin1999PhysRevB.59.443,Rehn16PRL,Rehn17PRL,Benton16NComms,Taillefumier2017PhysRevX.7.041057,Yan20PRL,Benton21PRL,davier2023combined,Gembe2023arxiv}.
	
	Despite their general instability to perturbations at absolute zero temperature ($T=0$), the substantial entropy that CSLs possess at low energies allows them to dominate the finite temperature physics in a finite region of model parameters. Moreover, CSLs often serve as `parent states' or an intermediate temperature regime for quantum spin liquids (QSLs), which arise when quantum fluctuations introduce dynamics among the classical ground states \cite{RVB_Fazekas,RK_1988,MS_RVB,Hermele2004PhysRevB.69.064404,Gingras2014RPPh,Sibille2018NatPh,Gaudet2019PhysRevLett.122.187201,Gao2019,Sibille2020,poree2023fractional}.

	Therefore, understanding and classifying spin liquids is of great importance. Successful classification schemes for quantum spin liquids  have been developed based on the projective symmetry group \cite{Wen2002PhysRevB.65.165113} and the modern perspective of gapped QSLs  \cite{essin-hermele2013,Barkeshli2019PRB}. 
	In contrast, the classification of CSLs has made much slower progress.
	Previous works have attempted to classify frustrated classical spin systems using constraint counting \cite{Moessner1998PhysRevB.58.12049}, linearization around given spin configurations \cite{Roychowdhury2018PRB}, supersymmetry-inspired constructions \cite{Roychowdhury-arXiv}, or identification of topological invariants tailored to specific lattices \cite{Benton21PRL}.
	
	Recently, a more general scheme has been proposed for classifying CSLs based on their energy spectrum \cite{Han2023arXivCLS1,Han2023arXivCLS2}.
	More concretely, the scheme utilizes the connection between CSLs and physics of flat bands at the bottom of the Hamiltonian's spectrum, responsible for the extensive degeneracy of the classical ground states ((Fig.~\ref{Fig_overview}(a)).
	In this classification scheme, CSLs are divided into two categories.
	The first is characterized by a topological invariant that persists as long as the lowest flat bands in its energy spectrum remain separated by a gap from the higher dispersive bands. 
	This category is called ``fragile topological'' classical spin liquids (FT-CSLs, Fig.~\ref{Fig_overview}(b)) since the topological characteristics can be made to disappear by adding spins to the unit cell without closing the spectral gap.
	The second category, called algebraic CSLs (Fig.~\ref{Fig_overview}(c)), occupies the boundaries between FT-CSLs where the spectral gap closes, as illustrated schematically in Fig.~\ref{Fig_overview}(d).
	The eigenvector configurations around the gap-closing points determine the emergent Gauss's law describing the algebraic spin correlations in this category of CSLs.
	
	In this work, we advance the above classification scheme by investigating the consequences of crystalline symmetry on the classification of classical spin liquids, in analogy to how crystalline symmetries enrich the classification of topological phases.
	Specifically, motivated by the classification of band representations in Topological Quantum Chemistry (TQC) \cite{bradlyn2017topological}, we develop a mathematical framework for  determining how the flat bands transform under symmetry, and elucidate whether symmetry protects the gap closing between the bottom flat bands and the higher dispersive bands. 
	If the gap closing is protected, then the symmetry forbids a FT-CSL, which is a significant constraint for model building or material analysis.
	Further details of the symmetry representations yield information on the topological classification of FT-CSLs and the degeneracy structure of the algebraic CSLs.

	Importantly, the present mathematical framework goes beyond how TQC is used to classify electron band structures.
	Specifically, the symmetry data of the lattice spins is 
	generally not sufficient to classify the CSL.
	An additional piece of information -- the emergent, virtual lattice of the constrainer terms in the Hamiltonian and their crystalline symmetry properties -- is crucial and must be incorporated into the symmetry analysis. 
	
	In the following, 
	we briefly review the  constrainer Hamiltonian formalism of CSLs and its classification in Sec.~\ref{SEC_review_classification}.
	We then give the recipe of the abstract crystalline symmetry analysis in Sec.~\ref{SEC_Crystalline_sym}. 
	This recipe is then applied to two known models on the Kagome model in Sec.~\ref{SEC_kagome_application} as a demonstration. 
	We then introduce a new pinch-line model with symmetry-protected nodal line degeneracies guided by our insight from the symmetry classification  in Sec.~\ref{SEC_new_model}.
	Finally we summarize our results in Sec.~\ref{SEC_summary}.

	\section{Brief review of  the constrainer Hamiltonian and CSL classification} 
	\label{SEC_review_classification}
	We study spin models in the limit of a large number of spin components $\mathcal{N}$.
	This is equivalent to adopting a ``soft spin'' approximation, where the constraint on the spin length $\mathbf{S}^2 = 1$ is enforced only on average as $\langle {S}^2 \rangle  = 1$, by introducing a Lagrange multiplier or ``chemical potential'' to the spins in the Self-Consistant Gaussian Approximation, a method generalized from the Luttinger-Tisza method~\cite{Luttinger1946PhysRev.70.954,Luttinger1951PhysRev.81.1015}.
	This approximation has been demonstrated to be valid for many Heisenberg candidate CSLs~\cite{Garanin1999PhysRevB.59.443,Conlon2010PRB, Conlon2009PRL,Isakov2004PhysRevLett.93.167204,Schmidt_2022}. 
	The Hamiltonians of these CSLs can be written in the \textit{constrainer  form}, 
	\[ 
	\label{eq:H_constrainer}
	\mathcal{H} =     \sum_{\mathbf{R} \in \text{u.c.}} \sum_{I=1}^M \, [\mathcal{C}_I (\mathbf{R} )]^2\  ,
	\] 
	where for a given constraint index $I$, 
	the \textit{constrainer} $\mathcal{C}_I(\mathbf{R})$ is a scalar that defines a linear combination of a local cluster of spins inside and around the unit cell located at $\mathbf{R}$ (see Eqs.~(\ref{eqn:kagomeAFM_2}), (\ref{eq:HRSM}) for concrete examples).
	The Hamiltonians we consider are translationally invariant and consist of sums of squared constrainers.
	
	A more explicit way to  express $\mathcal{C}_I(\mathbf{R})$ without referring to pictures of the lattice is  to write
	\[\label{EQN_abstract_Ham_general}
	\begin{split}
		\mathcal{H}   &= \sum_{\mathbf{R} \in \text{u.c.}}\sum_{I=1}^M   [\mathcal{C}_I(\mathbf{R} )]^2 
		\\
		&= \sum_{\mathbf{R} \in \text{u.c.}}\sum_{I=1}^M\left[
		\sum_{\mathbf{r}} \mathbf{S}(\mathbf{r})\cdot \mathbf{C}_I(\mathbf{R},\mathbf{r})
		\right]^2 \ .  
	\end{split}
	\]
	Here, $\mathbf{S}(\mathbf{r}) = (S_1, \dots, S_N) (\mathbf{r})$ is the vector whose $N$ components are the spins on the $N$ sublattice sites respectively.
	For example, $S_b(\mathbf{r})$ is the $b$-th sublattice site in the unit cell labelled by $\mathbf{r}$.
	The term $\sum_{\mathbf{r}} \mathbf{S}(\mathbf{r})\cdot \mathbf{C}_I(\mathbf{R,\mathbf{r}})$ is the constrainer $\mathcal{C}_I(\mathbf{R})$ written in a more explicit form, 
	that is, 
	\begin{align}
		&\mathbf{C}_I(\mathbf{0}, \mathbf{r}) = \left(\begin{array}{c}
			\sum_{j \in \text{1st sub-lat. sites   } } c_{1,j} \delta_{\mathbf{r}, \mathbf{a}_{1,j}} \\
			\sum_{j \in \text{2nd sub-lat. sites   } } c_{2,j} \delta_{\mathbf{r}, \mathbf{a}_{2,j}} \\
			\vdots\\
			\sum_{j \in \text{$N-$th sub-lat. sites   } } c_{N,j} \delta_{\mathbf{r}, \mathbf{a}_{N,j}} 
		\end{array}\right) \ ; \\
		&   \mathbf{C}_I(\mathbf{R},\mathbf{r}) = \mathbf{C}_I(\mathbf{0} ,\mathbf{r} -\mathbf{R}) 
	\end{align} 
	encodes exactly the information of how spins are summed together in $\mathcal{C}_I(\mathbf{R})$. The $c_{a,j}$ are numerical coefficients, and $\delta_{\mathbf{r}, \mathbf{a}_{a,j}}$ encodes the spin in the unit cell $\mathbf{a}_{a,j}$ on sublattice $a$, which is summed with coefficient $c_{a,j}$.
	
	Consider a system with $N$ sublattice sites and $M$ linearly independent constrainers per unit cell, where $M<N$. 
	There is an extensive degeneracy for the ground states in this system since the condition of all constrainers being zero does not fix all the spins. 
	Consequently, the Hamiltonian spectrum in momentum space has $N-M$ degenerate flat bands, each corresponding to one set of these degenerate ground states.
	Above these flat bands, there are $M$ higher dispersive bands that encode the finite-energy states.
	
	References \cite{Han2023arXivCLS1,Han2023arXivCLS2}  have discussed in great detail how the structure of the bottom flat bands and the higher dispersive bands can be used to classify the CSLs.
	For simplicity, let us use the case of $M=1$ (one constrainer per unit cell) to illustrate this scheme. 
	The scenario yields $N-1$ flat bands at zero energy, which corresponds to spin states obeying the constraint $\mathcal{C}(\mathbf{R}) = 0$; we have dropped the index $I$ since there is only one constrainer per unit cell.
	Additionally, there is a higher dispersive band describing spin states violating the constraint. 
	The eigenvector of this dispersive band, denoted as $\mathbf{T}(\mathbf{q})$, can be expressed analytically and is precisely the Fourier transform of the constrainer $\mathbf{C}_I(\mathbf{0}, \mathbf{r})$. 
	The dispersion of the higher band is  $\omega_T(\mathbf{q}) = |\mathbf{T}(\mathbf{q})|^2$  (see example in Sec.~\ref{SUBSEC_kagome_hex} in addition to the detailed formulation in  Refs.~\cite{Han2023arXivCLS1,Han2023arXivCLS2}).
	
	The  spectrum of the Hamiltonian encodes the classification of the CSL into one of two categories, determined by whether the dispersive band has a singular band touching point with the flat band or not.
	Within each category, a finer classification can be made by examining  the configuration of eigenvectors around the gapless point (first category) or its global topology  (second category). 
	In more detail, the classification is as follows.
	
	\noindent\textbf{1. Algebraic CSL:} If a gap closure point exists between the bottom flat bands and the higher dispersive band, the system is an algebraic CSL with algebraically decaying spin correlations. Here, the ground states conform to a charge-free Gauss's law, which is derived from the Taylor expansion of $\mathbf{T}(\mathbf{q})$ around the  band touching point. Specifically, if for the $a^\text{th}$ spin in the unit cell the lowest order term in the expansion is of order $m_a\geq1$ 
	\[
	T_a (\mathbf{q}) = \sum_{j=0}^{m_a} c_{aj} (-ik_x)^{j} (-ik_y)^{m_a-j}, \quad a = 1,\dots, N,
	\]
	then the ground states described by the spin configurations orthogonal to $\mathbf{T}  (\mathbf{q})$, i.e., 
	\[
	\mathbf{T}(\mathbf{q})  \cdot  \tilde{\mathbf{S}} (\mathbf{q}) = 0,
	\] in momentum space.
	Reverse-Fourier transforming this back to real space, we obtain the Gauss's law
	\[
	\sum_{a=1}^N\left(\sum_{j=0}^{m_a} c^\ast_{aj} (\partial_x)^{j} (\partial_y)^{m_a-j} S_a \right) \equiv \sum_{a=1}^ND_a^{(m_a)} S_a = 0,
	\]
	where $D_a^{(m_a)}$ denotes the differential operator of order $m_a\geq 1$ from Fourier transforming the $(-ik_x)^{j} (-ik_y)^{m_a-j}$ terms. This principle also applies to models with multiple constraints per unit cell (see Refs.~\cite{Han2023arXivCLS1,Han2023arXivCLS2} for detailed discussions). 
	
	\noindent\textbf{2. Fragile Topological CSL:} When the bottom flat band is entirely gapped from the higher dispersive band, $\mathbf{T}(\mathbf{q})$ becomes a non-zero, smoothly defined vector field in the target manifold $\CP^{N-1}$ (if complex) or $\RP^{N-1}$ (if real) across the entire BZ. 
	It can be classified by how it winds around the BZ, which is a $d$-torus, ${T}^d$. 
	The winding is encoded by the relative homotopy group $[T^d, \CP^{N-1}]$ (or $[T^d, \RP^{N-1}]$) of the map
	\[
	\hat{\mathbf{T}}(\mathbf{q}): {T}^d \rightarrow \CP^{N-1}  (\text{or }\RP^{N-1})
	\label{eq:map-T(q)}
	\]
	The homotopy class is invariant under smooth changes to the Hamiltonian as long as it maintains the constrainer form and the gap between the bottom flat and upper dispersive bands. 
	If the map $\hat{\mathbf{T}}(\mathbf{q})$ belongs to a nontrivial homotopy class, the corresponding gapped phase is (fragile) topological. Otherwise, the CSL is topologically trivial.  The fragility of the classification stems from the fact that adding (say, $P$) spins to the unit cell without closing the spectral gap changes the target manifold to $\CP^{N+P-1}$ (or $\RP^{N+P-1}$), whose relative homotopy group may be trivial.
	The homotopy class may also change by closing the spectral gap, at which point a band touching characterizes an algebraic CSL. 
	Thus, \textbf{the boundaries of fragile topological CSLs are algebraic CSLs.}
	The homotopy classification generalizes to systems with multiple degenerate flat bands or multiple higher bands (also see Refs.~\cite{Han2023arXivCLS1,Han2023arXivCLS2} for detailed discussions).

	\section{Crystalline symmetry analysis}
	\label{SEC_Crystalline_sym}

	\begin{table*}[th]
		\caption{Common lattices and their band representations ($\mathcal{BR}$)~\cite{bradlyn2017topological,elcoro2017double,vergniory2017graph,elcoro2021magnetic}; notation defined below Eq.~\eqref{eqn:BR_FB}.  
			\label{TABLE_lattices}
		} 
		\begin{tabular}{c @{\hskip 10pt} c c c } 
			\toprule
			Lattice & Space group $\cal G$ & $\mathcal{BR}$ of scalar/spin  & Representation at high symmetry points \\[3pt] \midrule 
			Lieb  & $P4/mmm$ & $({\mathrm A}_{\mathrm g})_{1{\mathrm a}}\uparrow {\cal G} \boxplus ({\mathrm A}_{\mathrm g})_{2{\mathrm f}}\uparrow {\cal G}$ &
			$(2\Gamma_1^+\oplus \Gamma_2^+) + (2X_1^+\oplus X_4^-) + (M_1^+\oplus M_5^-)$ \\ 
			square &$P4/mmm$ & $({\mathrm A}_{\mathrm g})_{1 {\mathrm c}} \uparrow {\cal G}$ & $(\Gamma_1^+)+(X_4^-)+(M_4^-)$\\
			kagome & $P6/mmm$ &$({\mathrm A}_{\mathrm g})_{3{\mathrm f}}\uparrow {\cal G}$ & ($\Gamma_1^+\oplus\Gamma_5^+)+(K_1\oplus K_5)+(M_1^+\oplus M_3^-\oplus M_4^-)$\\	
			hexagonal & $P6/mmm$ &$({\mathrm A}_{1\mathrm g})_{1{\mathrm a}}\uparrow {\cal G}$ & $(\Gamma_1^+)+(K_1)+(M_1^+)$ \\
			honeycomb & $P6/mmm$ & $({\mathrm A}_1')_{2{\mathrm c}}\uparrow {\cal G}$ & $(\Gamma_1^+ \oplus \Gamma_4^-)+(K_5)+(M_1^+\oplus M_4^-)$ \\
			pyrochlore & $Fd\bar 3m$ & $({\mathrm A}_{2\mathrm g})_{16{\mathrm c}}\uparrow {\cal G}$ & $(\Gamma_2^+\oplus \Gamma_4^+)+(X_2\oplus X_4)+(W_1\oplus W_2)$ \\ 
			\bottomrule
		\end{tabular}
	\end{table*}
	
	In previous studies~\cite{Han2023arXivCLS1,Han2023arXivCLS2}, several models utilizing the constrainer formalism have been reviewed and proposed (see Tables II and III in Ref.~\cite{Han2023arXivCLS2}).
	To classify a Hamiltonian written in the constrainer form, it is typically necessary to Fourier transform it into momentum space, 
	diagonalizing the Hamiltonian.
	
	An essential inquiry at this juncture concerns the ability to determine the class of the CSL solely based on the crystalline symmetry information of the model, without knowing the exact form of the constrainer. 
	The question is motivated by analogy to TQC~\cite{bradlyn2017topological}, which uses symmetry to constrain the topology and connectivity of band structures without knowledge of the exact Hamiltonian.
	The answer to this question provides significant insights into the physics of any lattice model, such as discerning whether an algebraic CSL is protected by crystalline symmetry, or merely accidental.

	In this section, we outline the symmetry analysis employed to identify band touchings and determine the topology of the flat bands within the constrainer Hamiltonian. As one may anticipate, this analysis has a close  relation to  the band irreducible representations  (irreps) of the crystalline symmetry group  used in TQC. 
	However, this approach by itself is insufficient to produce the CSL classification, as we will now explain.
	
	One notable example that will be extensively examined later in Sec.~\ref{SEC_kagome_application} is the comparison between the kagome antiferromagnetic (AFM) model and the kagome hexagon model. Although both models feature spins arranged on the kagome lattice, the former exhibits symmetry-protected gapless points on the bottom flat band, whereas the latter does not. 
	While the band symmetry analysis does predict gap-closures at specific high symmetry points within the Brillouin Zone (BZ), it fails to discern whether these closures occur on the bottom flat band(s) (which is crucial for the CSL classification) or among the higher dispersive bands (which is irrelevant for the CSL classification).
	
	From this shortcoming, we  discover that, apart from the irrep analysis of the microscopic spins and their lattice symmetries, the constrainers, whose centers define an ``auxiliary lattice'' that is generically distinct from the lattice of spins, play a critical role in determining the physics of the CSL. 
	This second aspect of physics goes beyond the irrep analysis of the local spins and encapsulates crucial information regarding the properties of the flat bands.
	
	We now describe the symmetry classification scheme.
	Consider a CSL consisting of spins on a lattice $S$, which is invariant under a space group $\mathcal{G}$.
	Every local spin at a particular lattice site transforms as a representation of the site-symmetry group at that site.
	The local representation induces a band representation $\mathcal{BR}_S$, which describes the symmetry of the entire spectrum, i.e., of both the dispersive bands and the flat bands~\cite{bradlyn2017topological,elcoro2017double,vergniory2017graph,elcoro2021magnetic,cano2018building}.
	While $\mathcal{BR}_S$ contains information about band touching points in the spectrum, it does not distinguish band touching points between the dispersive bands themselves from band touchings between the dispersive and flat bands. Thus, to derive symmetry constraints specific to the latter, we need more information.
	
	The extra information lies in the constrainers.
	Specifically, the dispersive bands live in the Hilbert subspace spanned by the constrainers. Thus, to distinguish properties of the flat and dispersive bands, we must apply TQC to the constrainers.
	Let $C$
	denote the lattice  comprised by the centers $\mathbf{R}$ of each constrainer ${\cal C}_I(\mathbf{R})$ in Eq.~\eqref{eq:H_constrainer}. Note that this lattice is virtual, in a sense that it is generically distinct from the lattice $C$ inhabited by physical spins.
	Since the constrainers must also satisfy the space group symmetry, each constrainer transforms as a representation of the site-symmetry group of the corresponding site in the lattice $C$, which induces a band representation ${\cal BR}_C$ describing the symmetry of the dispersive bands.
	
	Since the flat bands and dispersive bands together comprise the entire spectrum, the band representation of the flat bands is determined by
	\begin{equation}
		\label{eqn:BR_FB}
		\mathcal{B}_\text{FB} = \mathcal{BR}_S  \boxminus {\cal BR}_C  .
	\end{equation}
	where $\boxminus$ denotes the ``difference'' of two band representations, i.e., for each point in the BZ, the set of representations in ${\cal BR}_S$ not contained in ${\cal BR}_C$. 
	Examples of band representations are listed in Table~\ref{TABLE_lattices}, where $+$ indicates appending irreps at different momenta, $\oplus$ indicates the addition of irreps at the same momentum, and $\boxplus$ indicates the union of two band representations, defined analogously to $\boxminus$.

	The symmetry data determines the symmetry enforced band touching points and the topology of the gapped flat bands, as follows:
	\begin{itemize}
		\item If ${\cal BR}_C \nsubseteq  {\cal BR}_S$, then the spectrum has symmetry protected band touching points on the bottom flat band and the model is an algebraic CSL. 
		
		The degeneracy of symmetry protected band touching points between the flat bands and the dispersive bands are determined by the subduced representations of ${\cal BR}_C$ and ${\cal BR}_S$ at each momentum ${\mathbf q}$. Specifically, the number of dispersive bands is  dim$({\cal BR}_C\downarrow {\mathbf q} \,\cap\, {\cal BR}_S\downarrow {\mathbf q})$ and 
		similarly, the number of zero energy states $n_0$ can be deduced from the band representations (see Appendix~\ref{app:proof}).

		\item 
		If ${\cal BR}_C \subseteq  {\cal BR}_S$, then   the spectrum has either  no   band touching points on the bottom flat band, or the band touching points are not protected by symmetry. The system belongs to FT-CSL class.
		
		In this case,  at each $\mathbf{q}$ the irrep of the constrainer band is contained in ${\cal BR}_S$, and  the flat bands are fully gapped from the dispersive bands. 
		The topology of the flat bands is determined by its symmetry data vector $\mathcal{B}_\text{FB}$ and can be either atomic (trivial) or non-trivial but fragile, classified using the symmetry indicators developed for electron band structures ~\cite{song2020fragile,song2020twisted,fang2021filling,fang2021classification}.
	\end{itemize}
	
		The classification of flat bands in our constrainer Hamiltonians is mathematically the same as the classification of flat bands in the bipartite single-particle band structures studied in Ref~\cite{cualuguaru2022general}. In Appendix~\ref{app:bipartite} we prove their equivalence by constructing a map between the two systems, which relies on the introduction of the auxiliary lattice defined by the constrainers.

		\section{Application: two Kagome models}
		\label{SEC_kagome_application}
		We now demonstrate the symmetry analysis with a concrete example of the  kagome AFM model versus the kagome hexagon model (cf. Fig.~\ref{Fig_Kagome_AFM_lattice_proj}(a) for the kagome lattice).
		The two models have the same band representation ${\cal BR}_S$ because they both feature spins on the kagome lattice.
		However, their CSL properties are different. 
		The former is an algebraic CSL whose spectrum has a gapless point on the bottom flat bands, and the ground state fluctuations obey the emergent Maxwell Gauss's law $\bm{\nabla}\cdot \mathbf{E} = 0$. 
		The latter is a topological CSL, with two degenerate bottom flat bands that are fully gapped away from the higher dispersive band. Its higher dispersive band eigenvector $\mathbf{T}(\mathbf{q})$ is a three-component real vector on $S^2$, which has homotopy classes known as skyrmion number.
		The distinction between the two models can only be discerned by considering the symmetry of the constrainers.
		
		\subsection{Kagome AFM model}
				
				

			\begin{figure*}[ht!]
				\centering
				\includegraphics[width=0.9\textwidth]{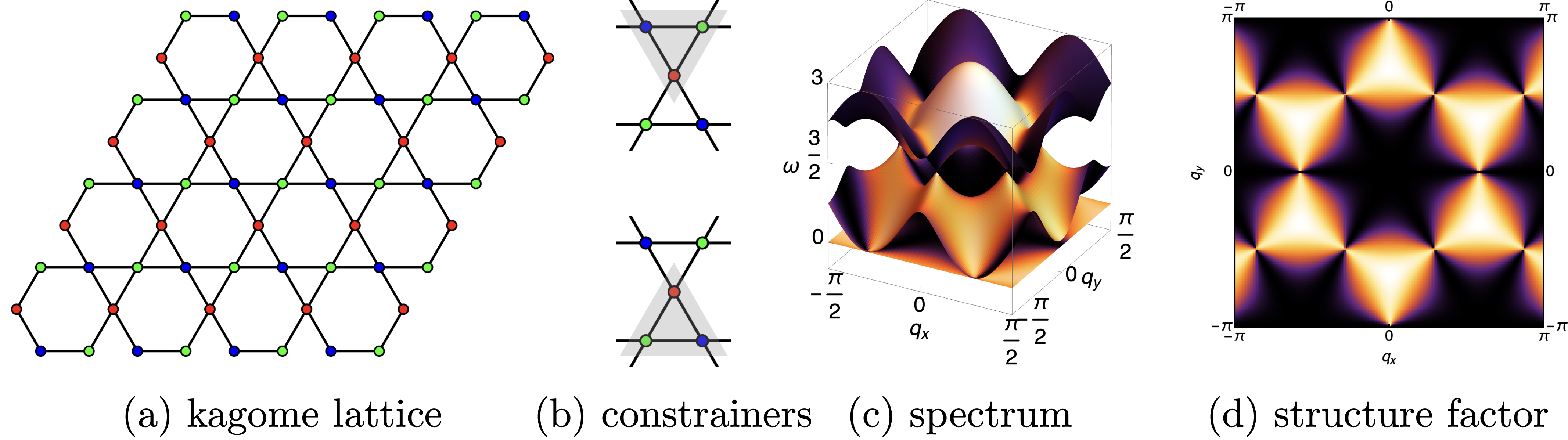}
				\caption{  (a) Kagome lattice for the Kagome AFM model (Eq.~\eqref{eqn:kagomeAFM_2}).
					(b) The two constrainers of the kagome model involve sites in the shaded regions. The ground states  are defined by the constraint that the sum of spins on each triangle must vanish (Eq.~(\ref{eqn:kagomeAFM_2})).
					(c) Spectrum $\omega({\bf q})$ that arises from diagonalizing
					the Hamiltonian Eq.~\eqref{eqn:kagomeAFM_2}. There is one  flat band  at the bottom of the spectrum and two higher dispersive  bands with gap-closing points on the flat band.
					(d) Spin structure factor showing pinch points at the position of gap-closing points.
				}
				\label{Fig_Kagome_AFM_lattice_proj}
			\end{figure*}

			The Hamiltonian of the kagome AFM model is 
			\[
			\label{eqn:kagomeAFM_2}
			\begin{split}
				\mathcal{H}_\mathsf{KAFM} & =  \sum_{\langle i,j\rangle} S_i S_j + 2 \sum_i S_i^2 \\ 
				& = \sum_\bigtriangleup\left( \sum_{i \in \bigtriangleup} S_i\right)^2 + \sum_\bigtriangledown\left( \sum_{i \in \bigtriangledown} S_i\right)^2 \\
				& \equiv  \sum_{\mathbf{R}  \text{ for } \bigtriangleup }
				[\calC_\mathsf{KAFM1}(\bfR)]^2 +\sum_{\mathbf{R} \text{ for } \bigtriangledown }
				[\calC_\mathsf{KAFM2}(\bfR)]^2 , 
			\end{split}
			\]
			The Hamiltonian consists of two constrainers $\calC_\mathsf{KAFM1,2}(\bfR)$, defined by sums over the spins on an up- or down-pointing triangle centered at $\bfR$, respectively (see Fig.~\ref{Fig_Kagome_AFM_lattice_proj}(b)); the sums in Eq.~\eqref{eqn:kagomeAFM_2} are over the constrainer centers. Ground states are hence defined by the  two  constraints:
			\[
			\calC_\mathsf{KAFM1,KAFM2}(\bfR) = 0  \quad  \forall \ \mathbf{R}\ ,
			\label{eq:kag_afm_constraint}
			\]
			on every triangular 
			plaquette.
			
			Diagonlizing the Hamiltonian in momentum space, we obtain the spectrum shown in Fig.~\ref{Fig_Kagome_AFM_lattice_proj}(c).
			There is one bottom flat band with gapless points where the dispersive bands touch.
			Using the techniques developed in Refs.~\cite{Han2023arXivCLS1,Han2023arXivCLS2}, expanding the dispersive band eigenvector at the gapless point yields the Gauss's law, 
			which turns out to be that of  Maxwell's theory:
			\[
			\begin{split}
				&3\partial_x (-  S_2 +   S_3)
				+ \sqrt{3} \partial_y (2  S_1 - S_2 -S_3 ) \\
				&\equiv \partial_x E_x + \partial_y E_y  = 0.
			\end{split}
			\]
			This is manifested in the pinch points in the equal-time spin correlation function shown in Fig.~\ref{Fig_Kagome_AFM_lattice_proj}(d).

			We now apply the symmetry analysis introduced in the previous section to prove that the band touching between the flat and dispersive bands is symmetry required.   
			This model is in space group ${\cal G} = P6/mmm$. The spins are located at the 3f Wyckoff position, which forms a kagome lattice. Each classical spin transforms as the scalar irrep ${\mathrm A}_{\mathrm g}$ of the site symmetry group. 
			Thus, 
			\[\label{EQN_BRL_kagome}
			{\cal BR}_S=({\mathrm A}_{\mathrm g})_{3{\mathrm f}}\uparrow {\cal G}.
			\]
			Its irreps at high symmetry points are listed in Table~\ref{TABLE_lattices}.
			
			The constrainers are centered around the 2c Wyckoff position, forming a honeycomb lattice where the honeycomb sites are located at the center of each triangle in the kagome lattice. Each constrainer transforms as the scalar irrep ${\mathrm A}_1'$ of the site symmetry group. Thus, 
			\[\label{EQN_BLbar_KagomeAFM}
			{\cal BR}_{C,\mathsf{KAFM}}=({\mathrm A}_1')_{2{\mathrm c}}\uparrow {\cal G}.\]
			Applying Eq.~(\ref{eqn:BR_FB}), we get the irreps of the flat band (see Table~\ref{TABLE_lattices})
			\[
			\begin{split}
				\mathcal{B}_{\mathrm{FB},\mathsf{KAFM}}& ={\cal BR}_S\boxminus {\cal BR}_{C,\mathsf{KAFM}} \\
				&=\left(\mathrm{A}_{\mathrm{g}}\right)_{3 \mathrm{f}} \uparrow \mathcal{G} \boxminus\left(\mathrm{A}_1^{\prime}\right)_{2 \mathrm{c}} \uparrow \mathcal{G} \\
				&=\left(\Gamma_5^{+} \ominus \Gamma_4^{-}\right)+\left(\mathrm{K}_1\right)+\left(\mathrm{M}_3^{-}\right)  .
			\end{split}
			\label{eq:BFBKAFM}
			\]
			At the $\Gamma$ point, 
			there is an irrep difference $\Gamma_5^{+} \ominus \Gamma_4^{-}$.
			The fact that an irrep difference appears rather than a sum indicates a band touching point at $\Gamma$ enforced by symmetry.

			\subsection{Kagome hexagon model}
			
			\label{SUBSEC_kagome_hex}
			\begin{figure*}[ht!]
				\centering
				\includegraphics[width=0.95\textwidth]{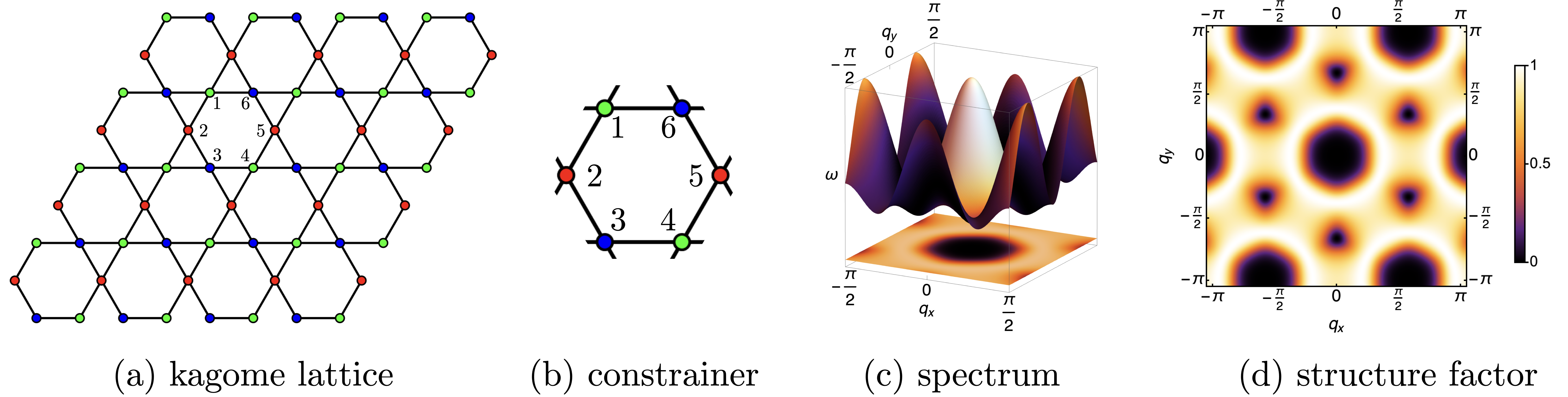}
				\caption{
					(a)  Kagome lattice for the Kagome hexagon model (Eq.~\eqref{eq:HRSM}).  
					(b) Constrainer of the Kagome-Hexagon model. Classical spins are arranged on a kagome lattice, with ground states defined by the constraint that the sum of spins on each hexagonal plaquette must vanish (Eqs.~\eqref{eq:RSM_M}).
					(c) Spectrum $\omega({\bf q})$ that arises from diagonalizing the Hamiltonian (Eq.~\eqref{eq:HRSM}) in momentum space. There are two degenerate flat bands at the bottom of the spectrum and a dispersive upper band with no band touchings between the upper and lower bands.
					(d) Spin structure factor showing an absence of singularities.
				}
				\label{Fig_RSM_all}
			\end{figure*}
			
			We now discuss the kagome-hexagon model \cite{Rehn17PRL} as an example of a fragile topological CSL with short-ranged correlations. 
			Its Hamiltonian is defined as 
			\begin{equation}
				\mathcal{H}_\mathsf{KH} = 
				\sum_{\mathbf{R} \in \text{all hexagons}} [\mathcal{C}_\mathsf{KH}(\mathbf{R})]^2\ ,
				\label{eq:HRSM}
			\end{equation}
			where the sum of $\mathbf{R}$ runs over hexagon centers on the kagome lattice
			(indicated in Fig.~\ref{Fig_RSM_all}(a,b)), or equivalently the centers of all unit cells.
			The constrainer 
			$\mathcal{C}_\mathsf{KH} (\mathbf{R})$ is the sum of the six spins around each hexagon  as labeled in Fig.~\ref{Fig_RSM_all}(a,b):
			\begin{eqnarray}
				\mathcal{C}_\mathsf{KH} (\mathbf{R}) =\sum_{i \in \text{hex. at }\mathbf{R}}   S_i\ .
				\label{eq:RSM_M}
			\end{eqnarray}
			The ground states are hence defined by the constraint 
			$ \mathcal{C}_\mathsf{KH}(\mathbf{R}) = 0    $ on every hexagonal plaquette.

			Diagonalizing $ \mathcal{H}_{\sf KH}$ in momentum space yields   a spectrum with three bands, of which the lowest two are flat and degenerate (Fig. \ref{Fig_RSM_all}(c)). 
			This is the case of one constrainer per unit cell discussed in Sec.~\ref{SEC_review_classification}, and 
			the eigenvector of the top band  can be found by Fourier-transforming the constrainer (as done in Refs.~\cite{Han2023arXivCLS1,Han2023arXivCLS2}),
			\[
			{\mathbf T} ({\mathbf q})
			= 
			\begin{pmatrix}
				\cos(\sqrt{3} q_x) \\
				\cos\left( -\frac{\sqrt{3}}{2} q_x + \frac{3}{2} q_y \right)  \\
				\cos\left( -\frac{\sqrt{3}}{2} q_x - \frac{3}{2} q_y \right)  
			\end{pmatrix}  ,
			\]
			and its dispersion is $\omega(\mathbf{q}) = \left|{\mathbf T} ({\mathbf q})\right|^2$.
			One can then explicitly see that 
			there are no band touchings between the upper band and the two flat bands at any point in the BZ, and consequently no pinch points in the correlation function (Fig. \ref{Fig_RSM_all}(d)).
			Accordingly, the real space correlations remain short ranged with a
			correlation length on the order of the nearest-neighbor distance at $T=0$.
			The ground state fluctuations are not described by any effective Gauss's law due to the absence of gapless points.
			
			We now use symmetry to explain why the flat band is gapped. 
			As in the previous example, this model is in space group ${\cal G} = P6/mmm$, and with spins are located at the 3f Wyckoff position, forming a kagome lattice. 
			The band representation is thus the same as in Eq.~\eqref{EQN_BRL_kagome}:
			\[{\cal BR}_S=({\mathrm A}_{\mathrm g})_{3 {\mathrm f}}\uparrow {\cal G}.\]

			The symmetry of the constrainers are different, however.
			They are located at the 1a Wyckoff positions, which form a triangular lattice with each site at the center of a hexagon in the kagome lattice. Each constrainer transforms as an irrep ${\mathrm A}_{1 {\mathrm g}}$ of the site symmetry group. Thus, their lattice irreps are
			\[{\cal BR}_{C,\mathsf{KH}}=({\mathrm A}_{1 {\mathrm g}})_{1 {\mathrm a}}\uparrow {\cal G},\]
			which is distinct from the previous example of the kagome AFM  (contrast with the Eq.~\eqref{EQN_BLbar_KagomeAFM}).
			
			Applying Eq.~(\ref{eqn:BR_FB}), we obtain (see Table~\ref{TABLE_lattices})
			\[
			\begin{split}
				\mathcal{B}_{\mathrm{FB},\mathsf{KH}}  & ={\cal BR}_S\boxminus{\cal BR}_{C,\mathsf{KH}} \\
				&=\left(\mathrm{A}_{\mathrm{g}}\right)_{3 \mathrm{f}} \uparrow \mathcal{G} \boxminus\left(\mathrm{A}_{1_{\mathrm{g}}}\right)_{1 \mathrm{a}} \uparrow \mathcal{G}\\
				&=\left(\Gamma_5^{+}\right)+\left(\mathrm{K}_5\right)+\left(\mathrm{M}_3^{-} \oplus \mathrm{M}_4^{-}\right) .
			\end{split}
			\label{eq:BFBKH}
			\]
			In contrast to Eq.~\eqref{eq:BFBKAFM}, no $\boxminus$ signs appear in the last line of Eq.~\eqref{eq:BFBKH}.
			Thus, all irreps of ${\cal BR}_{C,\mathsf{KH}}$ are included in ${\cal BR}_S$, which indicates that the flat band is fully gapped.

			However, the irreps that appear in $\mathcal{B}_{\mathrm{FB},\mathsf{KH}}$ 
			do not correspond to any sum of elementary band representations \cite{song2020fragile,fang2021filling,fang2021classification}. Thus, the flat bands have fragile topology, i.e., there is no way to understand them as coming from localized degrees of freedom.

			Comparing these two examples on the kagome lattice proves  that the irreps of the spins on the lattice, which determine ${\cal BR}_S$, do not provide enough information to determine the class of the CSL. 
			Specifically, ${\cal BR}_S$ requires a gap-closing at the $\Gamma$ point because of the two-dimensional irrep $\Gamma_5^+$,
			but it does not specify whether the band touching is between the flat band and a dispersive band (kagome AFM), between the two degenerate flat bands (kagome-hexagonal) or even between the two dispersive bands.
			To distinguish between these possibilities from symmetry requires the irreps of the constrainers, which are contained in  ${\cal BR}_C$.

			We note that the two constrainer models of spins on the kagome lattice discussed above
			can be mapped onto the two bipartite models of electrons discussed in the main text and Fig.~1 in Ref.~\cite{cualuguaru2022general}, with the distinction that the physical spins are situated on only one sublattice -- the other, virtual sublattice corresponds to the centers of the constrainers. The band representation analysis is however identical to that in Ref.~\cite{cualuguaru2022general}, as follows from the equivalence we prove in Appendix~\ref{app:bipartite}.

			\section{Construction of new classical spin liquid models}
			\label{SEC_new_model}
			
			\begin{figure*}[ht!]
				\centering
				\includegraphics[width=0.95\textwidth]{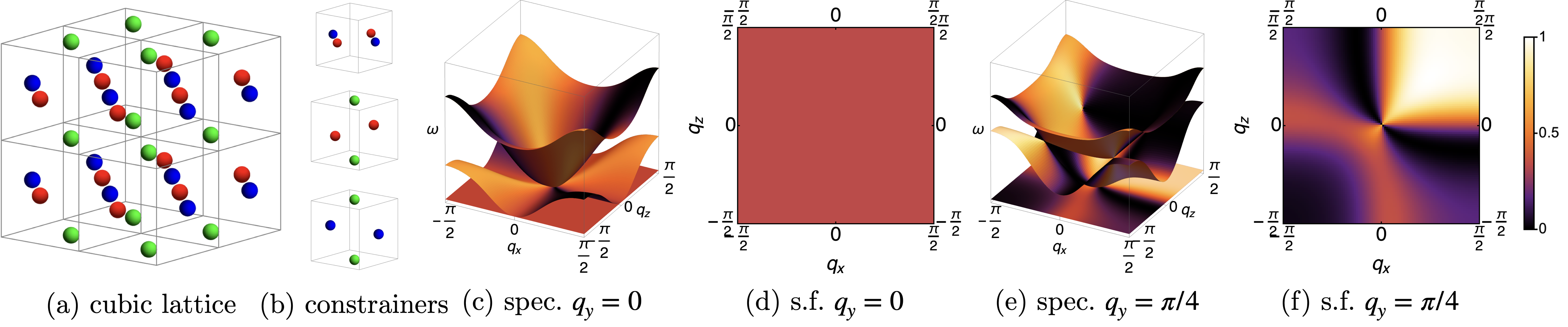}
				\caption{
					(a)  The cubic lattice. It has three sites per unit cell sitting on the face centers, forming three sublattices indicated in red, blue and green. 
					(b) Constrainers of the cubic  model. 
					(c) Spectrum $\omega({\bf q})$  from diagonalizing the Hamiltonian (Eq.~\eqref{EQN_PL_Ham}) in momentum space at $q_y = 0$. There is one flat band  at the bottom of the spectrum and  two higher dispersive bands, with band touchings  between the flat and dispersive bands along the lines $q_{x,z}= 0$.
					(d) Spin structure factor at $q_y = 0$.
					(e,f) Spectrum and spin structure factor at $q_y = \pi/4$.
				}
				\label{Fig_cubic}
			\end{figure*}

			\begin{table*}[ht!]
				\caption{The band representations ($\mathcal{BR}$) of the three-dimensional pinch line model~\cite{bradlyn2017topological,elcoro2017double,vergniory2017graph,elcoro2021magnetic}. 
					\label{TABLE_3dCSL}
				} 
				\begin{tabular}{c @{\hskip 10pt} c @{\hskip 10pt} c @{\hskip 10pt} c  } 
					\toprule
					&Spin on face center($S$) &Constrainer($C$) &Flat bands
					\\  [3pt]
					$\mathcal{BR}$ 
					&${\cal BR}_S=({\mathrm B}_{\mathrm 1})_{3{\mathrm d}}\uparrow {\cal G}$  
					&${\cal BR}_C=(^1{\mathrm E}_{\mathrm g})_{1{\mathrm a}}\uparrow {\cal G} \boxplus (^2{\mathrm E})_{1{\mathrm a}}\uparrow {\cal G}$  
					&${\cal BR}_S\boxminus{\cal BR}_C$ \\ [3pt] \midrule
					$\Gamma(0,0,0)$ &$\Gamma_4(3)$  &$\Gamma_2(1)\oplus\Gamma_3(1)$  
					&$\Gamma_4(3)\ominus \Gamma_2(1)\ominus \Gamma_3(1)$ \\
					$M(\pi,\pi,0)$ &$2M_1(1)\oplus M_2(1)$  &$2M_1(1)$  
					&$M_2(1)$ \\
					$R(\pi,\pi,\pi)$ &$R_1(1)\oplus R_2(1)\oplus R_3(1)$  &$R_2(1)\oplus R_3(1)$  
					&$R_1(1)$ \\
					$X(\pi,0,0)$ &$X_1(1)\oplus X_2X_3(2)$  &$2X_1(1)$  
					&$X_2X_3(2)\ominus X_1(1)$  \\
					$\Delta(u,0,0)$ &$\Delta_1(1)\oplus 2\Delta_2(1)$  &$2\Delta_1(1)$  
					&$2\Delta_2(1)\ominus \Delta_1(1)$ \\
					$Z(u,\pi,0)$ &$2Z_1(1)\oplus Z_2(1)$  &$2Z_1(1)$  
					&$Z_2(1)$ \\
					$T(u,\pi,\pi)$ &$3T_1(1)$  &$2T_1(1)$  
					&$T_1(1)$ \\
					$\Sigma(u,u,0)$ &$3\Sigma_1(1)$  &$2\Sigma_1(1)$  
					&$\Sigma_1(1)$ \\
					$\Lambda(u,u,u)$ &$\Lambda_1(1)\oplus\Lambda_2(1)\oplus\Lambda_3(1)$  &$\Lambda_2(1)\oplus\Lambda_3(1)$  
					&$\Lambda_1(1)$ \\
					\bottomrule
				\end{tabular}
			\end{table*}

			Let us now showcase the usefulness of the crystalline symmetry analysis by designing a new CSL model.
			It has symmetry protected \textit{nodal lines} on the bottom flat band, i.e. the gapless points form a line.
			In spin structure factor, the  nodal lines host pinch points around each point on the line, hence called \textit{pinch lines} in spin liquid literature \cite{Benton16NComms}.
			Although also in the algebraic CSL category, the pinch-line spin liquids exhibit very different physics than the more common algebraic CSLs with a gapless point. 
			It is thus interesting to construct robust pinch-line models protected by symmetry for future study.
			More specifically, we require the degeneracy nodal lines to be guaranteed to exist when both the number of constrainers and their symmetry property is fixed, but not the exact form of constrainers.
			Similar examples can also be found in Refs.~\cite{Benton16NComms,Han2023arXivCLS2}.


			We consider a model with a cubic unit cell and  
			spins located at the face centers (see Fig.~\ref{Fig_cubic}(a)). 
			In the large-$\mathcal{N}$ limit, these spins are effectively treated as scalars or soft spins.
			The orientation of each spin is pointing out of the face, in the positive $\hat{\mathbf{x}}, \hat{\mathbf{y}}, \hat{\mathbf{z}}$ directions respectively.
			For example, the   spin on the $x-$normal  face in points along $x-$direction.
			The model is symmetric under the symmetries of the magnetic space group ${\cal G} = P4'32'$ \cite{gallego2012magnetic,litvin2013magnetic} (No. 207.42 in the standard notation of Ref.~\cite{BNS-magnetic-space-groups}), which is generated by $C_{4,[001]}{\cal T}$, $C_{3,[111]}$ and $C_{2,[110]}$,  where the first subscript denotes the rotation order and the second three the rotation axis; $\mathcal{T}$ denotes time-reversal. Let us denote the spin on $x/y/z$-normal face as $S_{x/y/z}$ (note that they are not spin components but labels of spins on different sites). The symmetries act on the spins by:
			\begin{align}
				C_{3,[111]}:~& (S_x,S_y,S_z) \mapsto (S_z,S_x,S_y) \\
				C_{2,[110]}: & (S_x,S_y,S_z) \mapsto (S_y,S_x,S_z) \\
				C_{4,[001]}{\cal T}: & (S_x,S_y,S_z) \mapsto (S_y(-\mathbf a_x),-S_x,S_z),
			\end{align}
			where $S_y(-\mathbf a_x)$ indicates that the transformed spin on the $x$-normal face center of the current unit cell comes from the $S_y$ spin of the $-\mathbf a_x$ unit cell.
			
			We use the constrainers centered at the center of each cube (see Fig.~\ref{Fig_cubic}(b)):
			\begin{align}
				{\cal C}_{1}(\mathbf R) &= S_x(\mathbf R)-S_y(\mathbf R)-S_x(\mathbf R-\mathbf a_x)+S_y(\mathbf R-\mathbf a_y) \nonumber \\
				{\cal C}_{2}(\mathbf R) &= S_y(\mathbf R)-S_z(\mathbf R)-S_y(\mathbf R-\mathbf a_y)+S_z(\mathbf R-\mathbf a_z) \nonumber \\
				{\cal C}_{3}(\mathbf R) &= S_z(\mathbf R)-S_x(\mathbf R)-S_z(\mathbf R-\mathbf a_z)+S_x(\mathbf R-\mathbf a_x),
			\end{align}
			which obey the crystalline symmetries described in the preceding paragraph; the minus signs come from the transformation under $C_4{\cal T}$.
			The Hamiltonian is  then
			\[
			\label{EQN_PL_Ham}
			\mathcal{H}_\mathsf{PL} = \sum_{\mathbf R} \left [ {\cal C}^2_{1}(\mathbf R)+{\cal C}^2_{2}(\mathbf R)+{\cal C}^2_{3}(\mathbf R)\right ]
			\]
			
			The Fourier transformation  of the constrainers are
			\begin{align}
				\mathbf T_{1}(\mathbf q) &= 2i\left( \sin\frac{q_x}{2}, ~-\sin\frac{q_y}{2}, ~0 \right)^T
				\nonumber \\
				\mathbf T_{2}(\mathbf q) &= 2i\left( 0, ~\sin\frac{q_y}{2}, ~-\sin\frac{q_z}{2} \right)^T
				\nonumber \\
				\mathbf T_{3}(\mathbf q) &= 2i\left( -\sin\frac{q_x}{2}, ~0, ~\sin\frac{q_z}{2} \right)^T
				\label{EQN_3sCSL_Tq}
			\end{align}
			The three vectors span the subspace of the higher dispersive bands.
			Note that there are two dispersive bands and one flat band (see Figs.~\ref{Fig_cubic}(c,d)) since the identity $\mathbf T_{1}(\mathbf q)+\mathbf T_{2}(\mathbf q)+\mathbf T_{3}(\mathbf q)=0$ implies that the rank of the three vectors at a generic $\mathbf{q}$ is $2$.
			The line nodes are along the high-symmetry line $\Delta = (u,0,0)$ and its symmetry related partners in momentum space.

			The Gauss's law can be obtained by examining the eigenvector configuration around a gapless point. 
			At the $\Gamma$ point ($\mathbf{q} = \mathbf{0}$), the charge-free Gauss's law is
			\[
			\begin{split}
				\partial_x E_x - \partial_y E_y  = 0  ,\\
				\partial_y E_y -\partial_z E_z  = 0 , \\
				\partial_z E_z - \partial_x E_x  = 0   .
			\end{split}
			\]
			The two linearly independent constraints reflect the fact that both dispersive bands touch the bottom flat band at $\Gamma$.
			Along the line $\Delta = (0,u,0)$, the Gauss's law is 
			\[
			\partial_x E_x -\partial_z E_z  = 0 ,
			\]
			while the degree of freedom $E_y$ becomes ``gapped'' and is not involved in the Gauss's law.
			This is reflected in the spectrum  in Fig.~\ref{Fig_cubic}(c)
		where there is one higher dispersive band closes that touches the bottom flat band, and another stays gapped away from the origin.
		
		We now apply the symmetry analysis to determine the symmetry protected touchings between the dispersive bands and the flat bands. 
		The spins are located at the $3{\mathrm d}$ Wyckoff position of the aforementioned group $P4'32'$, corresponding to the face centers of the cubic unit cell. Each spin transforms as the irrep ${\mathrm B}_{1}$ of the site symmetry group $4'22'$~\cite{gallego2012magnetic,litvin2013magnetic}. 
		Thus, 
		\[\label{EQN_BRL_3dCSL}
		{\cal BR}_S=\left({\mathrm B}_{1}\right)_{3{\mathrm d}}\uparrow {\cal G}.
		\]
		Its irreps at high symmetry points are listed in Table~\ref{TABLE_3dCSL}.
		
		The constrainers are centered at the $1{\mathrm a}$ Wyckoff position which is at the center of a cube. The constrainers transform as the representation ${}^1{\mathrm E}\oplus {}^2{\mathrm E}$ of the site symmetry group $4'32'$. Thus, 
		\[\label{EQN_BLbar_3dCSL}
		{\cal BR}_C=\left({}^1{\mathrm E}\right)\uparrow {\cal G} \boxplus \left({}^2{\mathrm E}\right)\uparrow {\cal G}.\]
		Applying Eq.~(\ref{eqn:BR_FB}) and using the relevant band representations listed in Table~\ref{TABLE_3dCSL}, yields the representations in the flat band 
		\[
		\begin{split}
			\mathcal{B}_{\mathrm{FB}}& ={\cal BR}_S\boxminus{\cal BR}_C \\
			&=\left({\mathrm B}_{1}\right)_{3{\mathrm d}}\uparrow {\cal G} \boxminus \left({}^1{\mathrm E}\right)_{1{\mathrm a}}\uparrow {\cal G} \boxminus \left({}^2{\mathrm E}\right)_{1{\mathrm a}}\uparrow {\cal G}.
		\end{split}
		\]
		The representations at high symmetry points and along pertinent high symmetry lines are listed in Table~\ref{TABLE_3dCSL}.
		Importantly, Table~\ref{TABLE_3dCSL} shows that along the $\Delta$ line, there are irreps in $\mathcal{BR}_{C}$ which do not appear in ${\cal BR}_S$, guaranteeing that the dispersive band touches the flat band along the $\Delta$ line.

		\section{Summary}
		\label{SEC_summary}
		In summary, 
		we have developed the mathematical formalism for analysing the
		effects of crystalline symmetries on band touchings and topology of CSLs.
		In short, comparing the band representations generated by the constrainers to those generated by the spins determines whether the CSL is gapless (algebraic) or gapped.
		In the former case, the band representations also determine the degeneracy and location of gapless points in the BZ, while in the latter case, the symmetry data encodes the topology of the gapped band.
		

		Our symmetry analysis goes beyond applying TQC to the spins on the lattice: 
		on the contrary, as we have shown with explicit examples, understanding the symmetry of the constrainer terms in the Hamiltonian is imperative to deduce the band crossings. 
		This is different than the TQC classification of electron band structures, which holds independent of the form of the one-body Hamiltonian. We note that while the mathematical formalism of the `difference' of band representations in Eq.~\eqref{eqn:BR_FB} is identical to that developed for electron band structures on bipartite lattices in Ref.~\cite{cualuguaru2022general}, there are important distinctions. 
		First, the physical spins only inhabit one sublattice (S, which need not be bipartite), and there are no intersublattice `hopping' terms, which are central to the construction in Ref.~\cite{cualuguaru2022general}. 
		Second, the dual, virtual lattice (C) is determined solely by the centers of the constrainer clusters in Eq.~\eqref{eq:H_constrainer} and depends crucially on the form of the spin interactions. 
		In fact, the constrainer spin Hamiltonian can be thought of, in a sense, as a square of a one-particle hopping Hamiltonian (see Appendix~\ref{app:bipartite}).
		
		The formalism we have derived is a powerful tool for both understanding the robustness
		of spectral features in known models as well as for  reverse-engineering CSLs with desired properties.
		We have demonstrated the latter by introducing a new CSL with symmetry-protected nodal line degeneracies.
		The symmetry formalism will be an essential part of the comprehensive classification of the CLSs in the large-$\mathcal{N}$ limit going forward. 
		Our study paves way to a high-throughout examination of all possible lattice models that have specific spectral features of interest.

		
		\vspace{2mm}
		\section*{Acknowledgements} 
		H.Y. thank Owen Benton for helpful discussions. 
		H.Y. and A.H.N. were supported by the U.S. National Science Foundation Division of Materials Research under the Award DMR-1917511. 
		H.Y. and J.C. gratefully acknowledge support from the Simons Center for Geometry and Physics at Stony Brook University at which this project was initiated.
		Y.F. and J.C. acknowledge support from the National Science Foundation under Grant No. DMR-1942447.
		J.C. also acknowledges the support of the Flatiron Institute, a division of the Simons Foundation, and the Alfred P. Sloan Foundation through a Sloan Research Fellowship.
		
		\bibliography{reference}

		\onecolumngrid
		\appendix


		\section{The dimension of symmetry protected band touching}
		\label{app:proof}
		From topological quantum chemistry, given the orbitals on the lattice, the representation of bands in momentum space can be determined by the induced representation. In our case, the spin degrees on the lattice site form a representation $\rho_S$ of the site symmetry group, which is defined by the symmetries that leave the site $S$ invariant. Then the band representation of the full spectrum (flat bands and the dispersive bands) is given by ${\cal BR}_S = \rho_S\uparrow {\cal G}$. The corresponding representation of the little group at momentum $\mathbf q$ is ${\cal BR}_S \downarrow {\mathbf q}$. The explicit construction of these induced representations can be found in Refs.~\cite{bradlyn2017topological,cano2018building,cano2021band}.
		
		The representation of the constrainers can be determined in the same way. The constrainer $\mathcal C(\mathbf R)$ is a sum over a local cluster of spins as we have defined in Eq.~(\ref{eq:H_constrainer}). Denote its Fourier transform by ${\mathbf T}(\mathbf q)$. 
		If $\mathcal C(\mathbf R)$ transforms as the representation $\rho_C$ of its site symmetry group, then ${\mathbf T}(\mathbf q)$ transforms as a representation of the little group at $\mathbf q$ given by $\rho_C^{\mathbf q}=\rho_C \uparrow {\cal G} \downarrow {\mathbf q}$. The band representation of $\mathbf T(\mathbf q)$ is ${\cal BR}_C=\rho_C \uparrow {\cal G}$.
		Notice this is identical to the way to determine ${\cal BR}_C$ as constructed in the topological quantum chemistry~\cite{bradlyn2017topological,cano2018building,cano2021band}. However, the new feature that appears in this work is that the basis of the representation, ${\mathbf T}(\mathbf q)$, can vanish at a particular $\mathbf q$. When this happens, the transformation of $\mathbf{T}(\mathbf{q})$ under the little group symmetries is ill-defined, despite the fact that $\rho_C^\mathbf{q}$ is perfectly well-defined, since it is obtained from a Fourier transform of $\rho_C\uparrow \mathcal{G}$.
		Since $\mathbf{T}(\mathbf{q})$ vanishes precisely where the dispersive bands and flat bands touch, the symmetry data at this point gives information about the band touching, as we now describe.

		Since the Hilbert space of the dispersive bands are spanned by ${\mathbf T}(\mathbf q)$, the dispersive bands transform as the induced representation ${\cal BR}_C\downarrow {\mathbf q}$. On the other hand, the representation of the dispersive bands must be part of the complete representation induced from all spins in the model, ${\cal BR}_S\downarrow {\mathbf q}$.
		Let us now compare ${\cal BR}_S\downarrow {\mathbf q}$ to ${\cal BR}_C\downarrow {\mathbf q}$. There are two cases: (i) ${\cal BR}_C\downarrow {\mathbf q} \subseteq {\cal BR}_S\downarrow {\mathbf q}$. In this case there is no symmetry protected band touching between the dispersive bands and the flat bands since they transform as two representations ${\cal BR}_C\downarrow {\mathbf q}$ and ${\cal BR}_S\downarrow {\mathbf q} \ominus {\cal BR}_C\downarrow {\mathbf q}$. Note there may be accidental band touchings which are not protected by symmetry and hence not detected by this analysis.
		(ii) ${\cal BR}_C\downarrow {\mathbf q} \nsubseteq {\cal BR}_S\downarrow {\mathbf q}$.
		Such cases exist when ${\mathbf T}(\mathbf q) = 0$, where the induced representation ${\cal BR}_C=\rho_C\uparrow {\cal G}$ fails to describe the representation of the dispersive bands.
		Therefore,
		when an irrep in ${\cal BR}_C\downarrow {\mathbf q}$ does not appear in ${\cal BR}_S\downarrow {\mathbf q}$, it indicates a band touching. 
		
		Specifically, the dimension of the dispersive bands above zero energy is 
		\begin{equation}
			n_{\text{dispersive}} = \text{dim}({\cal BR}_C\downarrow {\mathbf q} \,\cap\, {\cal BR}_S\downarrow {\mathbf q})
		\end{equation}
		and the number of zero energy states $n_0$ is 
		\begin{align}
			n_0 &= \text{dim}\big({\cal BR}_S\downarrow {\mathbf q} \ominus ({\cal BR}_C\downarrow {\mathbf q} \,\cap\, {\cal BR}_S\downarrow {\mathbf q})\big)   \\
			&= \text{dim} \big( {\cal BR}_C\downarrow {\mathbf q} \,\cup\, {\cal BR}_S\downarrow {\mathbf q}) \ominus {\cal BR}_C\downarrow {\mathbf q} \big),
		\end{align}
		where $\ominus$ denotes taking the difference of irreps at each momentum. Away from the band touching point, $n_0$ gives the number of flat bands, while at the band touching point, $n_0$ gives the dimension of the band touching.

		\section{\label{app:bipartite} Mapping between the bipartite Hamiltonian and the constrainer Hamiltonian}
		Ref.~\cite{cualuguaru2022general} classifies flat electronic bands on bipartite lattices. In this Appendix, we show there is a bijective mapping between the constrainer Hamiltonians and the bipartite electronic Hamiltonians.
		
		The constrainer Hamiltonian of CSLs in Eq.~(\ref{eq:H_constrainer}) can be rewritten in terms of the $N\times 1$ constrainer vector ${\mathbf T}_I({\mathbf q})$, $I=1\dots M$, which is the Fourier transform of the constrainer $\mathcal C_I(\mathbf R)$ defined as the sum over a local cluster of spins, as follows
		\begin{equation}
			{\cal H} = \sum_{\mathbf q \in \text{BZ}} \sum_{i,j=1}^{N} s^\dagger_{\mathbf q,i} \big[ H_{\text{constrainer}}(\mathbf q) \big]_{ij} s_{\mathbf q,j}
		\end{equation}
		where the constrainer Hamiltonian matrix $H_{\text{constrainer}}$ is
		\begin{equation}
			\label{eqn:HTTeqHSS}
			H_{\text{constrainer}}(\mathbf q) =\sum_{i=1}^{M} {\mathbf T}_i({\mathbf q}) {\mathbf T}_i^\dagger({\mathbf q}) = \left( S^\dagger S \right)_{N\times N}
		\end{equation}
		by taking $S^\dagger=[{\mathbf T}_1({\mathbf q}),\dots,{\mathbf T}_{M}({\mathbf q})]$, and $s^\dagger_{\mathbf q,i}$ $s_{\mathbf q,i}$ are the creation/annihilation operators of our spin degrees at site $i$ and momentum $\mathbf q$.
		
		Now we construct an abstract block-off-diagonal matrix based on the matrix $S$ by
		\begin{equation}
			\label{eqn:Q}
			Q = 
			\begin{pmatrix}
				& S^\dagger_{N \times M}  \\
				S_{ M\times N} &
			\end{pmatrix}.
		\end{equation}
		Our constrainer Hamiltonian matrix appears in the first diagonal block of $Q^2$
		\begin{align}
			Q^2 &= \begin{pmatrix}
				H_{\text{constrainer}}(\mathbf q) & \\
				& \left( S S^\dagger \right)_{M\times M}
			\end{pmatrix}
		\end{align}
		We now define a set of auxiliary orbitals on the lattice sites formed by the constrainers, on which the creation/annihilation operators $c^\dagger_{\mathbf q,I}$ $c_{\mathbf q,I}$ are defined, where $I=1,\dots, M$. Then the matrix $Q^2$ represents a Hamiltonian over the Hilbert space spanned by the basis $(s^\dagger_{\mathbf q,i}, c^\dagger_{\mathbf q,I})$,
		where $i=1,\dots, N$ and $I=1,\dots, M$. 
		
		To connect with the bipartite Hamiltonian~\cite{cualuguaru2022general}, we interpret $Q$ as a Hamiltonian on a bipartite lattice where $S$ and $S^\dagger$ represent hopping between the two sublattices, which are denoted $L$ and $\bar{L}$ in Ref.~\cite{cualuguaru2022general}. However, in our CSL Hamiltonian, the bipartite lattice is virtual, i.e., it is a mathematical construction but does not correspond to a physical hopping between spins and constrainers. This construction shows that every constrainer Hamiltonian can be mapped to an electronic Hamiltonian on a bipartite lattice, and vice versa. 
		
		We now explain how to use the matrices $Q$ and $Q^2$ to re-derive the symmetry analysis we obtained in Appendix~\ref{app:proof}.
		First, the matrix $Q$ defined in Eq.~(\ref{eqn:Q}) has an anti-commuting chiral symmetry $\cal C$, satisfying $\{Q,{\cal C}\}=0$, given by
		\begin{equation}
			{\cal C} = 
			\begin{pmatrix}
				1_{N\times N}& \\
				& -1_{M \times M}
			\end{pmatrix}.
			\label{eq:C}
		\end{equation}
		%
		Thus, an eigenstate $\psi$ of the matrix $Q$ at energy $E$ has a chiral symmetry related partner $\cal C \psi$, which is an eigenstate of $Q$ at energy $-E$.  
		Both $\psi$ and ${\cal C} \psi$ are eigenstates of $Q^2$ with eigenvalue $E^2$. Since $[{\cal C}, Q^2]=0$, 
		one can construct linear combinations of these eigenstates that only live on the Hilbert subspace described by either the upper or lower block of $Q^2$, i.e., $\psi_{+} = \psi+{\cal C}\psi$ and $\psi_{-} = \psi-{\cal C}\psi$.
		These wavefunctions are defined in the basis of $(s^\dagger_{\mathbf q,i}, c^\dagger_{\mathbf q,I})$;
		projecting $\psi_\pm$ onto the upper/lower block yields an eigenstate of either $S^\dagger S$ or $S S^\dagger$.
		Thus, we have shown that the eigenvalue spectrum of $S^\dagger S$ and $SS^\dagger$ are identical, with one exception, which is where $\psi_\pm$ is ill-defined.
		Assuming $N>M$ and $\rank (S) = M$,
		this happens precisely when $\psi = \mathcal{C}\psi$, which implies $E=0$ and $\psi_-=0$.
		In other words, the constrainer Hamiltonian $S^\dagger S$ has zero energy eigenstates that are not eigenstates of $SS^\dagger$.

		Now consider the symmetry properties of the bands.
		In the constrainer models we consider, every symmetry $g$ commutes with $\mathcal{C}$, since symmetries do not mix spins with constrainers.
		Thus, if $\psi$ is simultaneously an eigenstate of $Q$ with energy $E$ and an eigenstate of a symmetry $g$ with eigenvalue $\xi$, then $\cal C \psi$, $\psi_{+}$ and $\psi_{-}$ are also eigenstates of $g$ with eigenvalue $\xi$ because $[{\cal C}, g]=0$. 
		It follows that the representation of the dispersive bands in $S^\dagger S$ and the bands of $SS^\dagger$ are also identical.
		However, the flat bands in $S^\dagger S$ do not have partners in $S S^\dagger$.
		Therefore the representation of the flat bands are given by the difference between the band representations of $S$ and $C$, i.e., ${\cal BR}_S \boxminus {\cal BR}_C$. 
		If the flat bands are gapped from the dispersive bands, then every irrep that appears in ${\cal BR}_C$ must also appear in ${\cal BR}_S$ since the dispersive bands of each have the same irreps. Thus, if an irrep appears in ${\cal BR}_C$ that is not in ${\cal BR}_S$, it must be that the flat bands are not gapped from the dispersive bands.
		In this case, the representation of the dispersive bands at the band touching point is not given by ${\cal {BR}}_{C}$ since the symmetry analysis only works for nonzero energy states. The results here are identical to what we have shown in Appendix~\ref{app:proof}.


	\end{document}